# Making Ideas at Scientific Fabrication Laboratories

Carlo Fonda and Enrique Canessa
*Science Dissemination Unit (SDU)*
*The Abdus Salam International Centre for Theoretical Physics (ICTP), 34151 Trieste, Italy*
email: scifablab@ictp.it

*Abstract:* Creativity, together with the making of ideas into fruition, is essential for progress. Today the evolution from an idea to its application can be facilitated by the implementation of Fabrication Laboratories, or FabLabs, having affordable digital tools for prototyping. FabLabs aiming at scientific research and invention are now starting to be established inside Universities and Research Centers. We review the setting up of the ICTP Scientific FabLab in Trieste, Italy, give concrete examples on the use in physics, and propose to replicate world-wide this class of multi-purpose workplaces within academia as a support for physics and math education and for community development.

1. **Overview**

Creativity and the conception of new thoughts, together with the action of bringing these ideas to reality, is essential for human growth and development. It is an ability that we all have and that we can cultivate with practice [1]. Furthermore, *"Scientific thought, and its creation, is the common and shared heritage of mankind"* –as quoted by Prof. A Salam, during the 1979 Nobel Prize in physics' ceremony [2]. Creative thinking needs to be nurtured and encouraged at all ages. The evolution from a creative expression to innovative applications for the benefit of the society, can today be facilitated by the implementation of small-scale Fabrication Laboratories, or FabLabs, offering (say, a more personal) digital fabrication facilities.

Learning happens in an authentic, engaging, personal context in which one goes through a cycle of imagination, design, prototyping, reflection, and iteration as one finds solutions to challenges or bring ideas to life. FabLabs, having affordable digital tools for prototyping [3-5], should start to be established also at Universities and Research Centers. There are two main reasons to encourage these novel initiatives. One is the fact that most of the logistics needed for FabLabs are already present in these places –like safety control and policies, adequate space and facilities, skilled personnel and financial support to run the structure and make the usage as free and open as possible. The second reason is the aim to foster the interaction between academics, students and the society –as an extra, and rich source for the input of new ideas to be developed together with professionals, playing the role of a more enriched FabLab environment for the benefit of all.

We briefly review here our experiences with the first year of activities within the Scientific Fabrication Laboratory (SciFabLab) of the ICTP in Trieste, Italy. We propose to replicate this class of multi-purpose workplace at Universities and Research Centers in developing countries –which is today feasible because of the unprecedented technological developments and the affordable costs involved. We discuss how SciFabLabs can be a open place to learn, and get notions of science, beyond the traditional classrooms and without limits.

## 2. The FabLabs

The concept of a FabLab was first imagined at the Center for Bits and Atoms (CBA) at the Media Lab of the Massachusetts Institute of Technology in USA, in 2001 [3]. FabLabs and Makerspaces are novel, small-scale workshop areas offering the possibility of digital fabrication and rapid prototyping to anyone interested. These are run mainly by enthusiasts called "Makers" and such structures form a distributed network of laboratories enabling invention and providing access to an array of flexible computer controlled tools that cover several different length scales and various materials. A FabLab is also a workplace for innovation, providing stimulus for entrepreneurship at local scales. A FabLab becomes a platform for learning and knowledge exchange: an open place to play, create, learn, mentor, share and invent. Fablabs are available as a community resource, offering open access to individuals as well as scheduled access to workshops and training activities in different technological subjects.

FabLabs include tools to laser cutting, to built 2D and 3D structures, CNC milling machines that make circuit boards and precision parts, low-cost 3D printers and scanners, and a suite of electronic components and programming tools for low-cost, high speed microcontrollers and tiny computers for electronic prototyping. These machines can produce prototypes with the aim to allow making "*almost anything*" and to prototype and refine new ideas. These prototypes embraces technology-enabled products generally perceived as limited to mass production and experimentation. Through Internet, Makers share experiences, results and processes adopting in most cases the open source model.

To define a laboratory as a FabLab, these criteria have to be met [4]:

- *Public access to the FabLab* is crucial since FabLab is all about democratization of tools and means for invention and personal expression.
- The FabLab has to *support and subscribe to the Fab Charter* –a set of rules describing what FabLab is, and what is not, and what are its duties and responsibilities.
- *FabLabs have to share a common set of tools and processes* to be able to produce the same thing in each and every FabLab in the world.
- *FabLabs must play an active part in the international FabLab network.*

## 3. Realization of Scientific FabLabs

Motivated by the academic activities that are of interest to scientists and by the formation of new communities interested in science and development around FabLabs, a Fabrication Laboratory inserted in a scientific framework can open new dimensions to science and education, inspire curiosity and offer powerful new ways to facilitate the development of new ideas with a certain impact [6]. With the small economical investment requested to establish a SciFabLab, an academic institution, specially in developing countries, can catalyze innovation and entrepreneurship at the grassroots level with the help of its faculty and students.

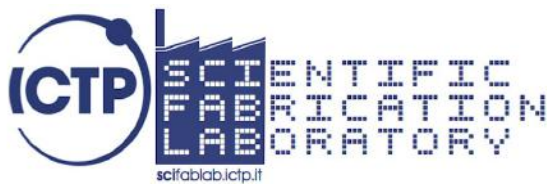

To date, the approximated costs (in dollars) of a small SciFabLab are for 3D printers, scanners: 2 - 5 K$ ; laser cutter: 5 - 10 K$ (the most expensive single equipment, but very useful); computer-controlled milling machine: 2 - 5 K$; electronics and microcontrollers: 2 K$ ; computers: not many are needed, most can be old PCs: 2 K$ ; other (non digital ;-) tools: 2 K$; consumables: 2 - 5 K$ / year ; plus personnel costs and other infrastructural costs, maintenance , etc.

The main requirement for making ideas at a SciFabLab is that free membership must be granted with an approved project in the fields of science, education and sustainable development [7], which stands for the period of the project itself. Through an ongoing *"Open Call for Guest Projects"*, a variety of projects are then fabricated, often spanning across multiple iterations. At the ICTP SciFabLab, a few examples so far developed have been on: thesis projects; robotics and automation, electronics, micro-controllers and computer–based science projects, 3D printing for education, green technologies and alternative energies, Smart Cities and Internet of Things applications. Also, there have been projects on innovative use of mobile devices to promote education and Mobile Science [8].

The staff members of a SciFabLab called *"Coordinators",* are in charge of management and coordination. They are assisted by some *"Managers"* to whom the management is delegated during the evenings' and weekends' opening hours of the SciFabLab. Makers carry out their *"Guest Projects"* and become *"Hosts"*. They are invited to accept some regulations (a kind of extended FabLab Charter). Acceptance of a Guest Project does not grant any reimbursement of costs incurred by the project team (such as purchase of material, deployment expenses, travel, or other), which have to be covered by the Hosts themselves. The SciFabLab workplace given to a *Guest Project,* and Internet connectivity, are given free of charge. Hosts are allowed to use machines, tools and facilities at the SciFabLab with the authorization of the Coordinator or FabLab Manager present on site. For some machines/tools, it is mandatory to follow a free technical and/or safety training course and to demonstrate the required skills before using them.

Designs and processes developed within a SciFabLab can be protected and sold whenever an inventor chooses to do so, but the project should remain available for individuals to use (for example, in home fabrication [9]) and to learn from. Commercial activities can be prototyped and incubated, but they must not conflict with other uses, they should grow beyond, rather than within, the SciFabLab. They are expected to benefit the inventors, Labs, and networks that contributed to their success.

Accepted members of a Scientific FabLab agree:
- to share their expertise with other members of the SciFabLab;
- to publish openly their results in the literature, on-line, or exhibit them in Maker events. The authors of any prototype, model or code can retain their full ownership, however it is highly recommended to release prototypes, models, and codes developed within the Scientific FabLab as open source/open hardware.

Since by "making" one can learn most effectively –*i.e., carrying out a physical implementation;* a Scientific FabLab can play a significant role as a hub to support the creative work of scientists and targeted scholar audiences. This interaction between academics, students and the society has an unprecedented development.

Particular attention is to be paid to original activities, projects, ideas and contents directed to developing countries, which can be reproduced at any other Scientific Fabrication Laboratory. By adopting new technologies in new ways, and sharing openly the results, SciFabLabs have many potentials for education and development. It can also help to inspire curiosity and understanding in young scholars and future generations of scientists.

4. Some Experiences at ICTP SciFabLab

The ICTP SciFabLab, inaugurated in August 2014, is equipped with low-cost, state-of-the-art and versatile computer-controlled rapid prototyping tools such as 3D printers, 3D scanners, laser engraving and cutting machines, and also open source software. It aims to provide a safe working platform for invention, creativity and resourcefulness within an international scientific environment. It is part of the global FabLab network [5] and at the same time, it is strongly rooted locally. The ICTP SciFabLab offers free access through proposed projects to a large, regional community of Makers with free access at regular opening hours in evenings and the weekend.

- The ScifabLab Makers need to be very strongly rooted locally, and actively work to be locally relevant.
- At the same time, they need to be networked globally, to cater easy knowledge exchanges and build upon experiments and designs from elsewhere in order to create a local impact.

The ICTP SciFabLab hosts many digital fabrication workshops to building prototypes. Scientists, visitors and staff of the ICTP also form part of this creative workplace of about 350 square meters. Every single project being developed is openly discussed via an associated WebBlog [5], which is updated by the Makers owners of the idea or the prototype to keep track of the progress of the projects.

In the past year, the SciFabLab offered 700 hours of opening time for Makers to work on their projects. These activities were also supported from the Comune of Trieste (Municipality) for allowing to pay two additional Manager assistants. Since the beginning (August 2014) we had more than 30 projects hosted in the SciFabLab. The fraction of rejected projects was in general low since before accepting projects the Makers were invited to discuss their ideas beforehand and to revise their proposed proposals. On average 5 - 10 users present when the FabLab is open to the public (48h/month, 100d/year), scientists during all workdays, also on mornings A network of >50 external makers were connected with us (15 were women) . ICTP visiting scientists have worked here for 2+ months each (Colombia, Cameroon, Nigeria) . We hosted 8 researchers from Developing Countries for 2-4 weeks and 6 students from University of Trieste and Udine did their thesis/working stage .

- **3D Dissemination**

Making mathematical functions tangible provides a new dimension for science dissemination. One goal to use low-cost 3D printers within SciFabLabs is to produce (or print) mathematical objects such as those realized for the IMAGINARY Open Mathematics Exhibition [10] that would otherwise be difficult to visualize. The shape of such objects can be seen in a more intuitive manner which helps to improve the understanding of certain complex functions in the fields of physics and mathematics. For example, Makers at SciFabLab have already worked out how to produce 17 special mathematical objects by 3D printing of filaments made of biodegradable plastic PLA (Polylactic acid) –an environmental-friendly material derived from corn starch. The final goal is to support museums, schools and higher education institutions in countries with poor scientific infrastructure. Also, the idea is to inspire curiosity, incentive deeper understanding and put learning literally in the hands of scholars.

- **Printed Objects from Recycled Plastic**

Rapid prototyping by 3D printers can also have a different style. In can also be used to teach students the basic principles behind the Fuse Deposition Modeling (FDM) 3D technology and to carry out research to produce, for example, conductive plastic filaments in order to be able to print circuits. In collaboration with condensed matter physicists from the Interdisciplinary Research Group GruMoc at the University of Cartagena there is some research in this context for using ionic elements inside the plastic filaments [11]. But before printing these conductive filaments, researchers needs to find a scalable solution to produce low-cost plastic filaments from bottle caps. The commercial options available which involve the use of plastic pellets are usually not cost-effective. Therefore, by developing an alternative easy and cheap way to produce plastic filaments within SciFabLabs, the next step is to make it conductive with a polymer matrix that will contain ions that can be transported electrically.

The applications for such conductive filaments can be numerous [12]. They can be used to build sensors, integrated circuitry and for the production of low cost wearable materials that can pass on signals and data for health control. Work along these lines is in progress. This is a research activity and not a production one in constant development.

- **Maker Expo of Open Technologies**

The gathering of Makers in a free Maker Expo is a unique occasion to exchange ideas between a lot of creative Makers (*i.e.,* people generating new tools and original solutions), scientists and scholars (*i.e.,* people that need to use new tools to solve problems). These public activities organized by the ICTP SciFabLab aim to share the learning and research that have been acquired during the making of something new using science and technology. These unique Expo activities attract lots of interest by the media and the people, and they are good platforms to broaden the knowledge about an institution.

One example is the organization of a "Mini Maker Faire" (under License Make Media Inc.). The need for organizing these structured, exhibition experiences is to offer an alternative learning appeal within an educational environment. Attendees stay thus connected with science and innovation, and they are encouraged to nurture curiosity and the "Maker spirit".

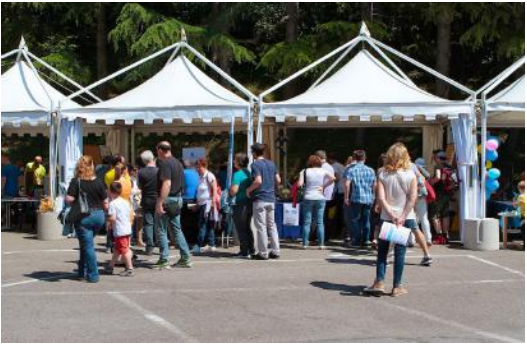
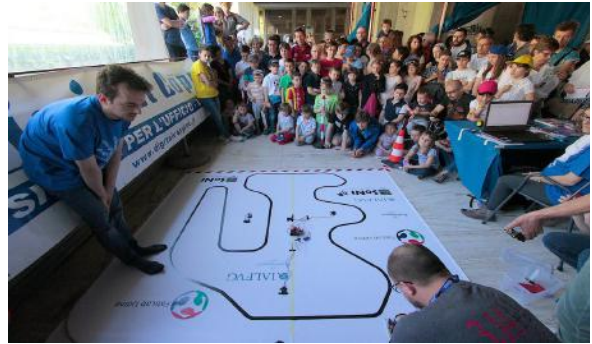

*Snapshots of the Trieste Mini Maker Faire 2016 (images M. Goina @ictpnews) -more at www.makerfairetrieste.it*

- **Few ICTP SciFabLab Prototypes in Pictures**

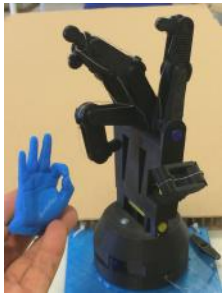

*First prototype for a "Voice-Controlled Artificial Handspeak System" –an automated translator for communication with the deaf to help reduce language barriers [13].*

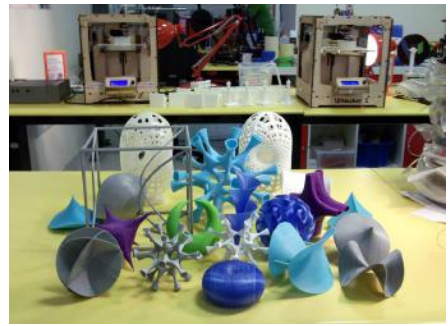

*3D printing output of the IMAGINARY Open Mathematics Sculptures [10], image by @scifablab.*

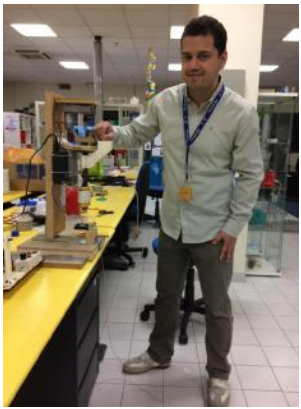
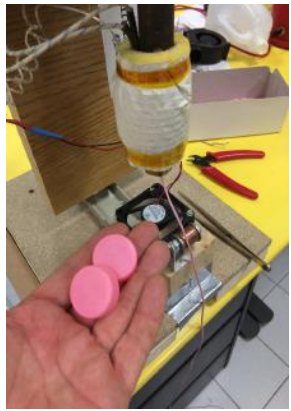
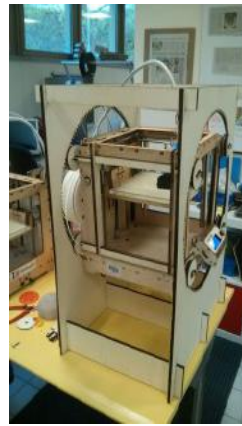
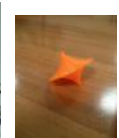
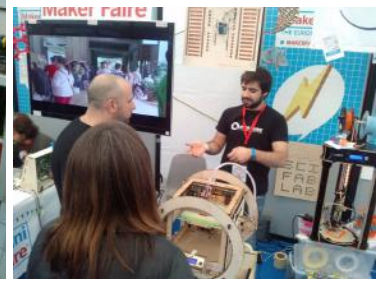

*The making of 3D Printer filament from recycled plastic [11] (images by @scifablab).*

*Prototype for inverted 3D printing of complex objects saving plastics and improving quality (images by @scifablab).*

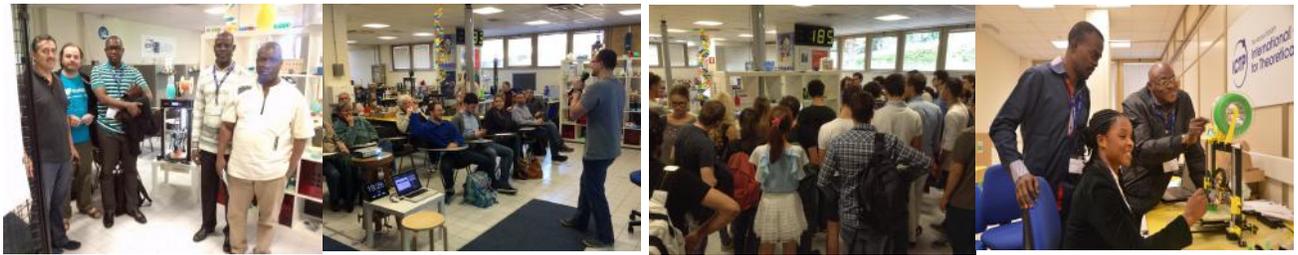

*Snapshots of visitors, workshops, conferences, hands-on activities at the ICTP SciFabLab (images by @scifablab).*

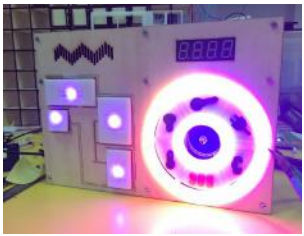

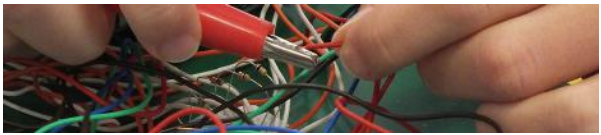

*"Hands(H)ome":* Prototype of home automation user interface for monitoring and control designed for elders and impaired people *(image by @scifablab).*

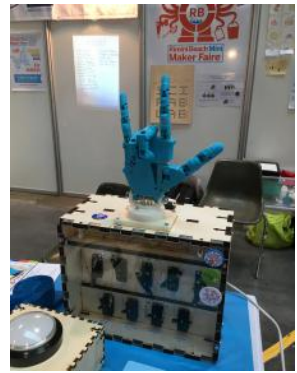

*"MANIpolare per comunicare"*: Design and creation of a low-cost Game Kit used as a learning support for sign language for children who are deaf and deaf-blind *(image by @scifablab).*

## 5. Do-It-Yourself (DIY) Physics Education

Newer generations of learners need to stay competitive in a growing digital society by being creative and innovative. Within a Scientific FabLab the goal is to help to cultivate and grow these skills, specially now that it is noticed that the undergraduate University curriculum lends itself to diminishing creativity in students [14] In academia, the emphasis in the first years is still usually on theory and mathematical models as opposed to a more modern approach based on practice, actual design and build process which can start already within a community space offering public, shared access to high-end manufacturing equipment. To this end, there exist a close relationship between the maker movement, maker spaces and FabLabs. In fact, *"everyone likes making things"* [15] be this inside a maker space (i.e., craft area) or a FabLab (workshop space with a set of electronics equipment).

SciFabLabs can be used to learn and get notions of science beyond the traditional context of books, chalkboards of even computer representations. SciFabLabs can contribute to physics education by means of hand-ons prototyping, the discovery and the physical representation of concepts and natural laws. In such workshop spaces, students can engage in their own projects and develop new skills by accessing the equipment needed for their prototypes and modeling. Since these open FabLabs are relatively new, their full effect and impact on the education process cannot be fully quantified yet [14]. However one main benefit of having a SciFabLab is that they certainly can offer a more open environment than a classroom which can be used more openly, without classroom scheduling constraints. In this regard some evidence already shows the benefits of establishing workshop spaces at Universities. As discussed in [14], the central pillar of maker spaces is the actual act of building and making. For example, at three different Universities (Washington, Pennsylvania State and Puerto Rico), the use of maker technologies has largely influenced a change in the curriculum for design and manufacturing toward what is known as the Learning Factory model. We foresee similar promising results for the implementation of taylored Scientific Fablab around the world. The Department of Electronic and Electrical Engineering of Obafemi Aolowo University in Ile-Ife, in Nigeria has started to integrate access to their recently created SciFabLab into courseware delivery [16]. Students are encouraged to use the available FabLab facilities in conducting their personal and departmental projects. Four courses with a total of 8 units now require the use of their FabLab facilities (two "Group Design" courses, as well as students' final year projects).

The informal learning processes that occur within FabLabs is done by manufacturing prototypes and refining 3D physical objects which are shared with a large community of scholars via Internet --as open source and/or open hardware technologies. This opens an effective connection between abstracted scientific and concrete applications as for example when assembling a 3D printer from scratch. Some physics concepts that are empirically learn by FabLab participants include thermodynamics, phase transitions, fluid dynamics, viscosity and electricity and magnetism to mention a few. The knowledge of the multidisciplinary field of material sciences and remote sensing becomes also important within a ScifabLab. For example the activities that participants engage in a SciFabLab connect to physics education via the discovery of the laws of thermodynamics. The fundamental physical concepts of temperature, equilibrium, conservation of energy, etc, are all experienced and explained under various circumstances when using the 3D printers. Their calibration and tunning processes are driven by the real-time measurements and control of temperature for different plastic materials being fused. The transition between the solid and viscous phases (glass transition) of the coloured filaments need to be controlled upon heating to its transition point, which usually have different values for every provider of the material. Notions of phase transitions, visual fluid flow, and properties of materials of different substances (conductive or not) are helpful in the processes to carry out practical prototyping work. The resistance of gradual stress (or pulling) of 3D printing filaments with a diameter of 1,75 mm (or 3 mm) extruded through a tiny small nozzle hole helps to explain the concept of viscosity as a function of temperature. In addition, the concepts of electric fields, electric current, etc, as well as the associated laws of electromagnetic induction and magnetism, play significant roles when programming and practicing with embedded devices and sensors. All of these physics concepts are learn, experienced and directly applied in the SciFabLab prototypes.

By enhancing hands-on learning, and interactive class activities, we put the process of learning literally into the hands of our maker students and scientists. For example by producing prosthetics like customized legs, arms, hands, etc we teach scholars about the human body and its kinetics. Students in a FabLab can also print 3D objects based on equations to better visualize complex structures. Physics abstraction can benefit from 3D printing technology's capability to reproduce and make malleable objects, shortening the time taken to discover alternative solutions and then evaluate quickly any possible new outcomes. Physics students can learn faster through practice and be more productive through experience for example by studying a complex Calabi-Yau manifold out of the plastic, which yield applications in superstring theory, or from a non-trivial, infinitely connected gyroid structure, that make them potential photonic crystals (see Figures below).

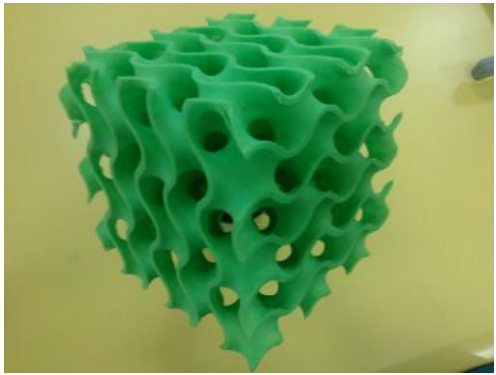

*"G*yroid structure" for photonic crystals
*(image by @scifablab).*

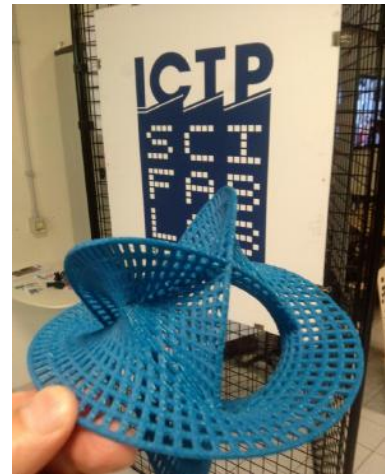

*"Calabi-Yau manifold out of the plastic"* which yield applications in superstring theory *(image by @scifablab).*

With the growing online design libraries, and the use of FabLabs facilities, anyone can today start building its own lab equipment at remarkably low-cost [17]. SciFablabs may play also a critical role in boosting entrepreneurship skills and the possibility of the marketing of new ideas, something that is completely missing in universities and scientific institutions –specially so in many places from developing countries.

Real physics can be learnt and incorporated in the 3D prototyping elements and not just by freemind designing to then printing. One of such, is the study of the dynamics of Moineau's Progressive Cavity pump coupled co-axially to an Auger screw (as shown in the picture below). This kind of complex systems found many applications in the automatization of food and oil industries and also, *e.g.*, in the research of the influence of suspended particles on the transition to turbulence in pure fluids [18]. A second example of the application of physics laws is the interactive visualization and understanding of nodal patterns on vibrating plates (Chladni figures), with the peculiarity of replacing sand with larger spheres moving on a 3D printed net of different geometries and under frequencies adjusted via an Arduino (see Figure from the project #PodobaZvoka by T. Druscovich).

Another good use in physics –also shown below, is the realization of a "Picosat" model. This is a light, small cube having solar cells, battery, GPS, compass, a small transceiver and a CPU that could be sent into space as ballast weight during the launch of bigger satellites. One field of application is the observation of large extensions of vegetation to establish its state of conservation by a photometer which could measure temporal color changes of the land. A working prototype of such photometer has been developed in the SciFabLab by M. Maris and his students for educational purposes.

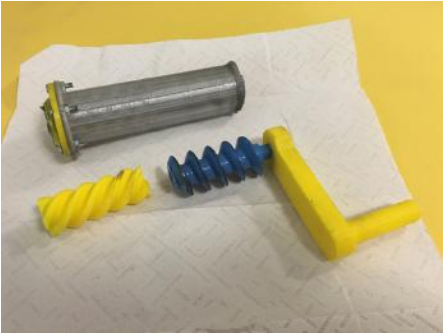
Moineau+Auger pump for volumetric flow proportional to the rotation rate
*(image by @scifablab)*

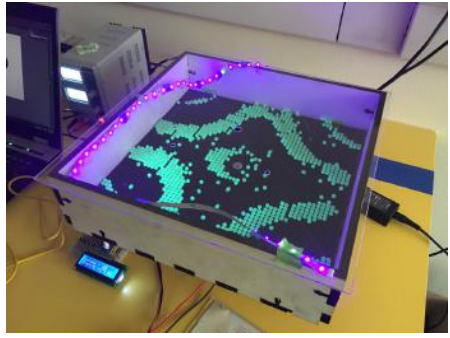
Chladni figures, with the peculiarity of replacing sand with larger spheres moving on a 3D printed net of different geometries and under frequencies adjusted via an Arduino
*(image by M. Goina @ictpnews)*

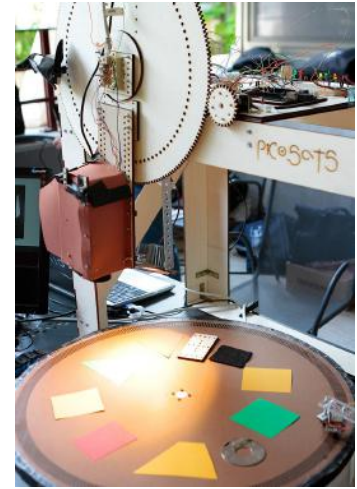
"Picosat" photometer model as ballast weight during the launch of bigger satellites
*(image by M. Goina @ictpnews)*

### 6. Fabricating the Future and Relevance for Developing Countries

In FabLabs one can learn by making objects and gain deeper knowledge about science, machines, materials and modeling from the hard work that goes into invention and innovation. These workplaces are growing fast in Europe and the USA, particularly at schools for hands-on STEM education. There are a few FabLabs also present in developing countries [19], which demonstrate the importance of this phenomenon.

On the other hand, it is foreseen that project-based SciFabLabs, as a FabLab space with special focus on possible scientific applications, will also start to grow [20,21]. An effective development of SciFabLabs relies on the awareness and clear understanding of the affordable digital technologies available and its implication for science education via a DIY method. As far as we know, besides the ICTP SciFabLab, there is also a Scientific FabLab being formed at different Universities: in Belgrade, Servia, for its different faculties [20], Nigeria [16,22] and Colombia [11].

The goal for creating SciFabLabs is to share knowledge and collaborate across international borders in the fields of science, education and sustainable growth [8,21]. Whereas the life in a high-skill manufacturing SciFabLab is built upon thoughts and creation.

The affordable and new digital technologies found in Scientific FabLabs can have an important bearing on physics teaching around the world as never happens before and can provide both low cost apparatus that are not commercially available or produce locally educational items in a easier way. Scientific FabLabs can enable researchers, students and makers in general to expand their work in new ways and to test out their ideas without breaking their budgets. These tools can also help new scholars to do science and discover it [3]. Creativity, together with the making of ideas into fruition, is essential for progress and can help build better communities.

*Acknowledgements*

Our sincere thanks go to Prof. Fernando Quevedo, ICTP Director, for his support in the establishment of the ICTP SciFabLab and to Ms. Margherita di Giovannantonio for the secretarial work behind the running of the SciFabLab's activities.